\documentclass[10pt]{iopart}

\usepackage{epsfig}
\begin{document}

\title[Strange Particle  Production in  p+p, p+Pb and Pb+Pb Interactions from NA49]{Strange
Particle Production
in  p+p, p+Pb and Pb+Pb Interactions from NA49}

\author{K. Kadija for the NA49
Collaboration \footnote[2]{For a full author list of the  NA49
Collaboration se reference \cite{mischke2001}}}  

\address{Rudjer Boskovic Institute, Zagreb, Croatia and \\
 CERN, Geneve, Switzerland}
\ead{Kreso.Kadija@cern.ch}

\begin{abstract}
 Recent  NA49 results on $\Lambda$, $\bar{\Lambda}$, $\Xi^{-}$ and
$\bar{\Xi}^{+}$ production  in
minimum bias p+p and centrality selected p+Pb  collisions at 158 GeV/c, and  
the results on $\Lambda$, $\bar{\Lambda}$, $K^{+}$ and $K^{-}$ production in central Pb+Pb collisions at
40, 80 and 158 A$\cdot$GeV are discussed and compared  with other available data. 
By comparing the energy dependence of $\Lambda$ and  $\bar{\Lambda}$ production
at mid-rapidity a striking similarity is observed between p+p and A+A data. 
This is also seen in the energy dependence of the $\Lambda$/$\pi$ ratio.
$K^{+}$/$\pi$ at mid-rapidity is affected in a similar way, due to the 
associated production of $K^{+}$ together with $\Lambda$ particles.
The observed  yields increase faster than the number of wounded
nucleons  when comparing p+Pb to p+p.
As already observed in A+A collisions, the increase is larger for
multistrange than for strange baryons and  for baryons than for
anti-baryons.
\end{abstract}




\section{Introduction}
The NA49 experiment \cite{na4999} offers a  unique opportunity  of collecting $-$ with the same detector $-$ data
on the full set of available hadronic interactions. The processes studied range from hadron-nucleon 
and centrality selected hadron-nucleus interactions to nucleus-nucleus collisions covering a wide range 
of energies and system sizes.

 In the context of this conference preliminary results concerning  strange and multistrange baryon  production in p+p, centrality selected
p+Pb at 158 GeV/c \cite{tanja2001}, as well as the results on strange baryon \cite{mischke2001} and meson
\cite{kollegger2001} production
from central Pb+Pb collisions at 40, 80 and 158 A$\cdot$GeV, will be presented. A critical discussion - including whenever possible
results from other
experiments - will attempt to
establish common features and differences between elementary and more complex hadronic collisions. 
The importance of maintaining a broad view on the totality of hadronic processes for the correct 
interpretation of the observables in specific systems will be emphasized.  
 
\section{Results and discussion}

$\Lambda$ and $\bar{\Lambda}$ particles are a good measure of the net-baryon 
density at mid-rapidity due to their charge and isospin neutrality.
An observed ratio  $\frac{\bar{\Lambda}}{\Lambda}$ $\ll$ 1 would imply that the
central region is still net-baryon dominated. This is due to baryon stopping, as 
refering to the transport of baryon number away from the nuclear fragmentation regions. 
On the quark level the net-baryon transfer from the
projectile and target fragmentation regions to
mid-rapidity, can
generally be described as a transfer of  net-quarks (to be
distinguished from the quarks  produced in  quark-antiquark pairs).
  From this point of view, the basic mechanism of
net-baryon transfer in p+p, p+A  and A+A collisions could  be very similar,
emphasizing  the necessity of a more detailed study  of particle production  in p+p and p+A
 collisions.

 The energy dependence of the
$\Lambda$ and $\bar{\Lambda}$ multiplicities  at mid-rapidity in  p+p
\cite{kadijapp}  collisions is shown in Fig.~1a. The averaged yields of
$\Lambda$ and $\bar{\Lambda}$ particles from $p$+$\bar{p}$
collisions at $\sqrt{s}$$=$200 and  $\sqrt{s}$$=$900 GeV are shown to indicate
the expected $\Lambda$ and $\bar{\Lambda}$ yields at higher energy. 
\begin{figure}[]
\vspace{-1.cm}
\epsfig{file=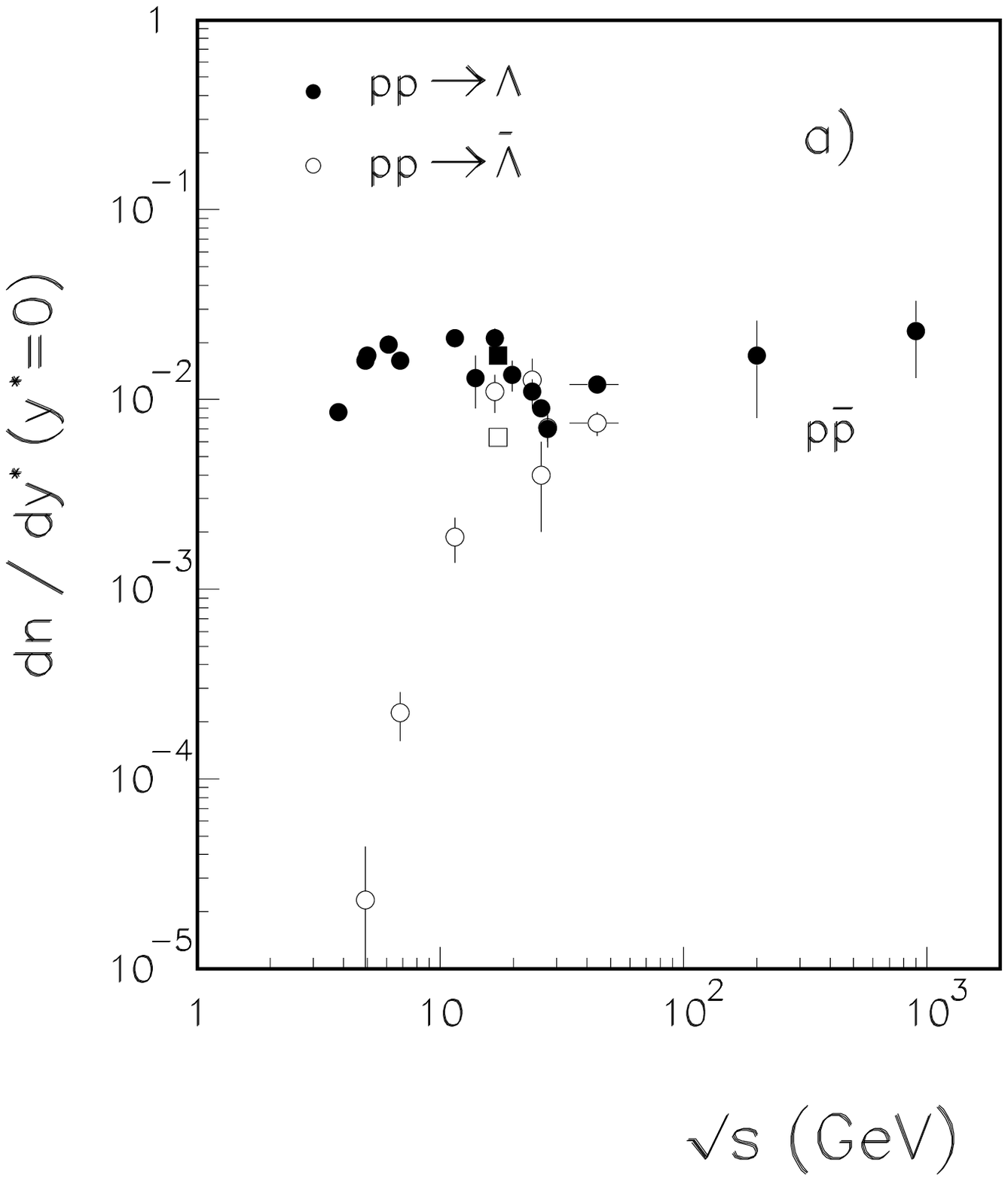,width=8.0cm,height=8.0cm}
\hspace{-2.0cm}
\epsfig{file=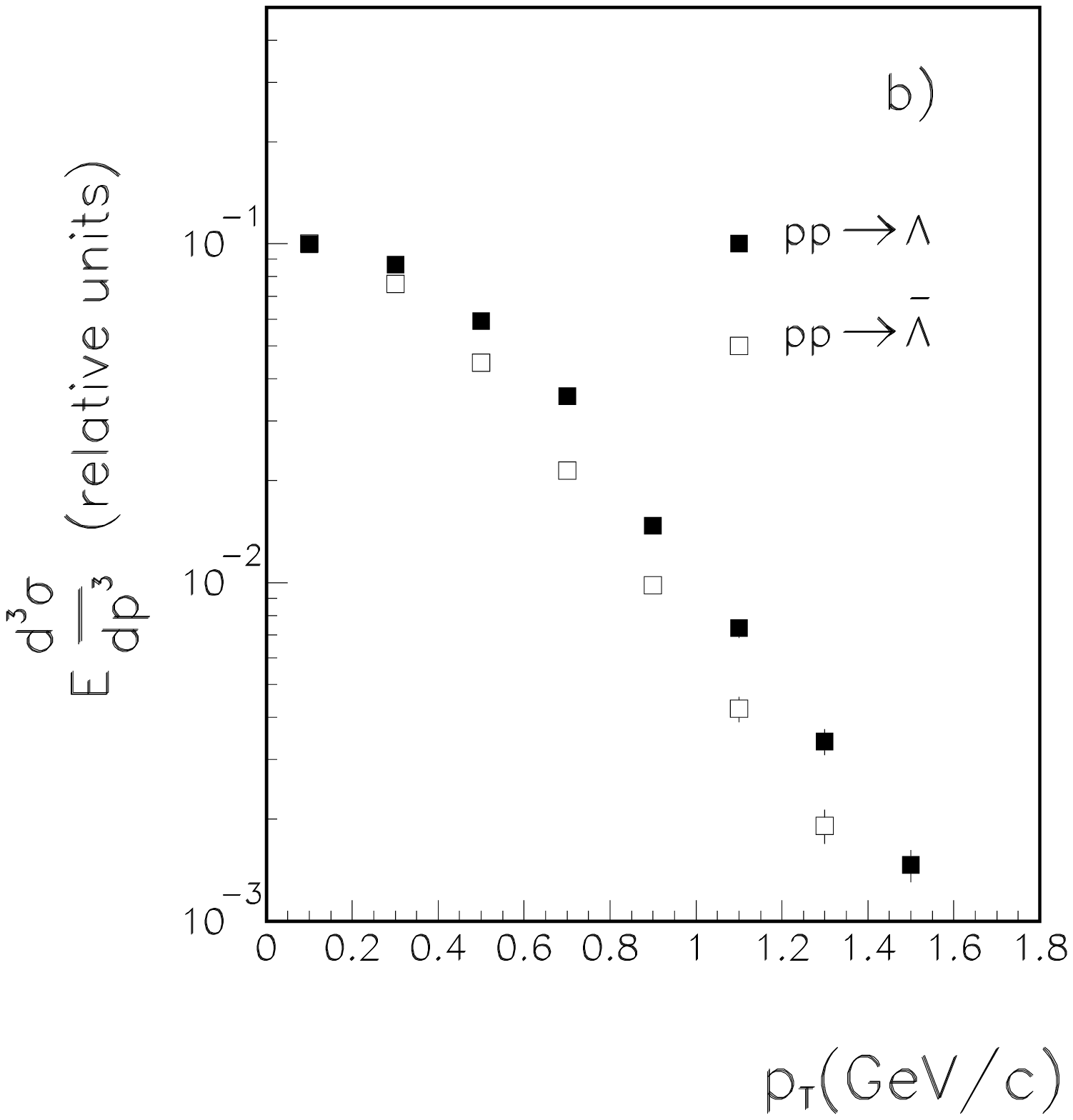,width=8.0cm,height=8.0cm}
\caption{a) Energy dependence of  $\Lambda$ and $\bar{\Lambda}$
mid-rapidity
multiplicities, including preliminary NA49 data (shown by squares), b)  Normalized $p_T$
distributions of
$\Lambda$ and $\bar{\Lambda}$ at mid-rapidity at 158 GeV/c (NA49     
preliminary).}
\label{fig:toosmall}
\end{figure}   
The  p+p data  for $\bar{\Lambda}$ shows a continuous rise with
nucleon-nucleon center-of-mass energy, $\sqrt{s}$.
Due to the isospin arguments the number of $\bar{\Lambda}$s measures
the number of pair-produced  $\Lambda$s.  
  The data is not conclusive enough, but there is
an indication that pair-produced $\Lambda$s  dominate
the mid-rapidity yield at energies above $\sqrt{s}$$=$20$-$30 GeV, as a
consequence of the very low net-baryon density in this kinematic region.
  In contrast to the $\frac{\bar{\Lambda}}{\Lambda}$ ratio, which is 
expected to be 1 if the zero net-baryon density is reached, the ratios of  
particles which are not charge and isospin neutral (e.g. $\frac{\bar{p}}{p}$,
$\frac{\bar{K^{-}}}{K^{+}}$) are expected to be less than 1. This is due to
the asymmetric pair production: for example $p$$\bar{n}$, $K^{+}$$\bar{K^{0}}$, etc. 

The  high net-baryon density at mid-rapidity at lower energies causes the
$\Lambda$ 
particles to be singly produced, essentially through
the associated production mechanism together with $K^{+}$ mesons.
 Available  results on
singly
produced  $\Lambda$s suggest  a very steep rise  after passing the
threshold energy.  At higher
$\sqrt{s}$, where the  pair-produced lambdas start to contribute
dominantly,
the yield of singly  produced $\Lambda$s is steeply
falling. The sum of these  two contributions shows a minimum  arround
$\sqrt{s}$$=$ 25 GeV, as a border between the regions of high and low
net-baryon density (see Fig.~1a).

 Preliminary NA49 data \cite{tanja2001} (squares in Fig.~1a)  shows that  approximately 40\% of the
$\Lambda$s at mid-rapidity  are pair-produced, and  60\% 
singly-produced, in associated production with $K^{+}$.

 Different particle $p_T$ spectra are expected for different production mechanisms. This 
correlation is demonstrated 
in Fig.~1b which  shows the normalized $p_T$ distributions of
$\Lambda$ and $\bar{\Lambda}$ hyperons measured at mid-rapidity.
 The steeper $p_T$  slope  observed for the $\bar{\Lambda}$s (pair produced
$\Lambda$s) than for the $\Lambda$s
 (singly and pair produced $\Lambda$s)
suggests that the $p_T$ spectra clearly depend on the particle  
production
mechanism. An interesting question is how this matches the physics
assumptions of the statistical model, regardless of its success 
in describing particle yields from p+p interactions.
\begin{figure}[]
\vspace{-1.cm}
\epsfig{file=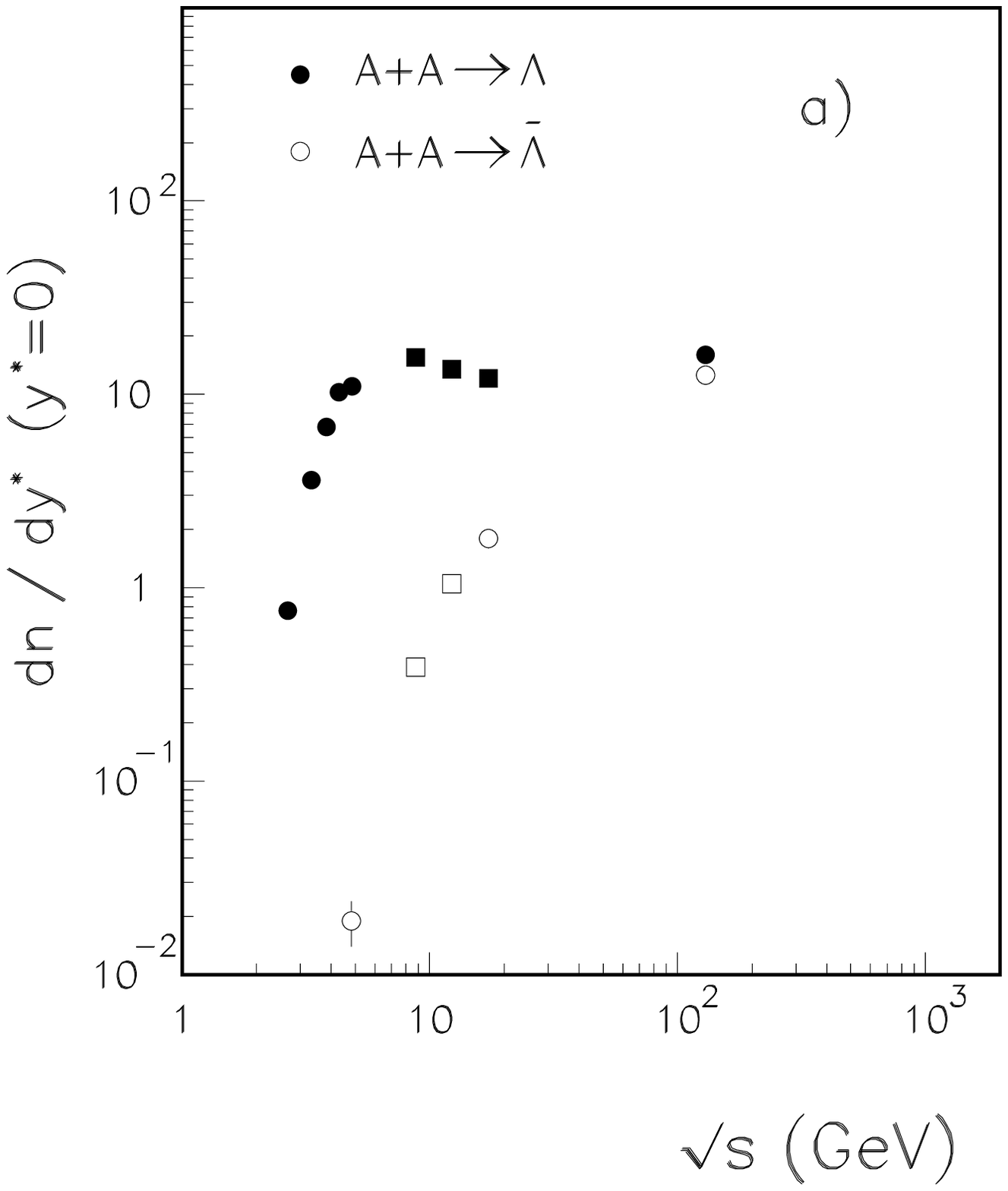,width=8.0cm,height=8.0cm}
\hspace{-2.0cm}
\epsfig{file=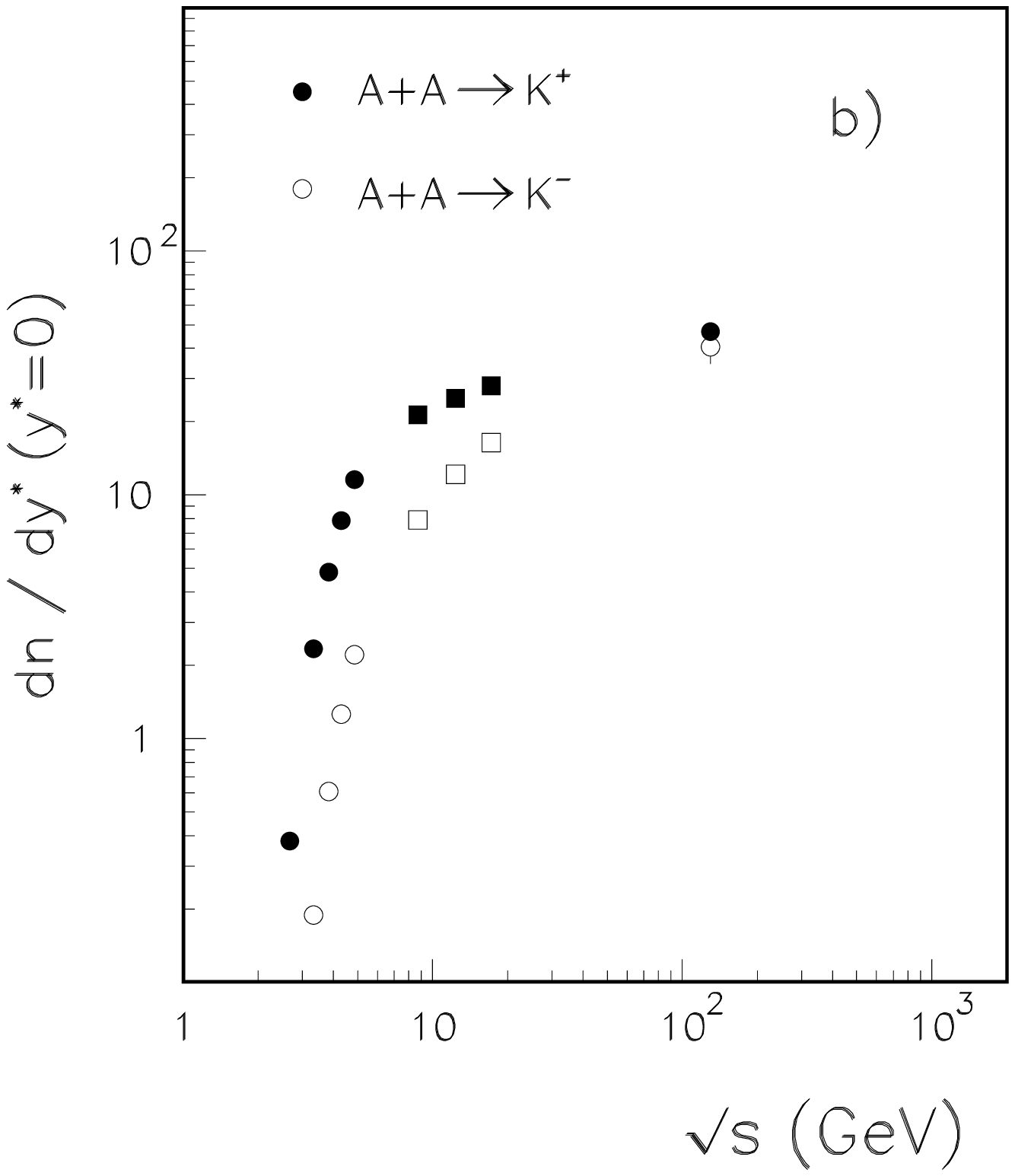,width=8.0cm,height=8.0cm}
\caption{ a) Energy dependence of  $\Lambda$ and $\bar{\Lambda}$,  
and  b) $K^{+}$ and $K^{-}$          
mid-rapidity
multiplicities in central A+A collisons. Preliminary NA49 data are
indicated by the squares.}
\label{fig:toosmall}
\end{figure}
Results from A+A collisions are ploted 
in Fig.~2a. Preliminary NA49  data \cite{mischke2001} on  $\Lambda$ and
$\bar{\Lambda}$
production at mid-rapidity (squares)  are compared to  other 
experiments \cite{aalambda} \cite{star2001} (circles).
Although at the same energy  A+A data shows a
higher degree of stopping than p+p data,
it is interesting to note a striking similarity in the energy
dependence of these two data sets. Both samples show
that  $\Lambda$ production at mid-rapidity is strongly correlated with
the change with $\sqrt{s}$  of the net-baryon density. For a  better understanding 
of the physical origin of the  net-baryon transfer and its correlation to particle
production, the precise p+p and A+A RHIC data, from the lowest up to top accesible energies,  
are needed.  It is not excluded that the  mid-rapidity $\Lambda$  yield in  A+A collisions will show a
minimum, like it was observed in p+p.
 A different origin  of this minimum is also not excluded. In p+p this could  indicate 
  the energy  region with (almost)  zero net-baryon density, and in
A+A  the region where the  net-baryon density at mid-rapidity starts to saturate 
towards a non-zero value.
\begin{figure}[]
\vspace{-1.cm}
\epsfig{file=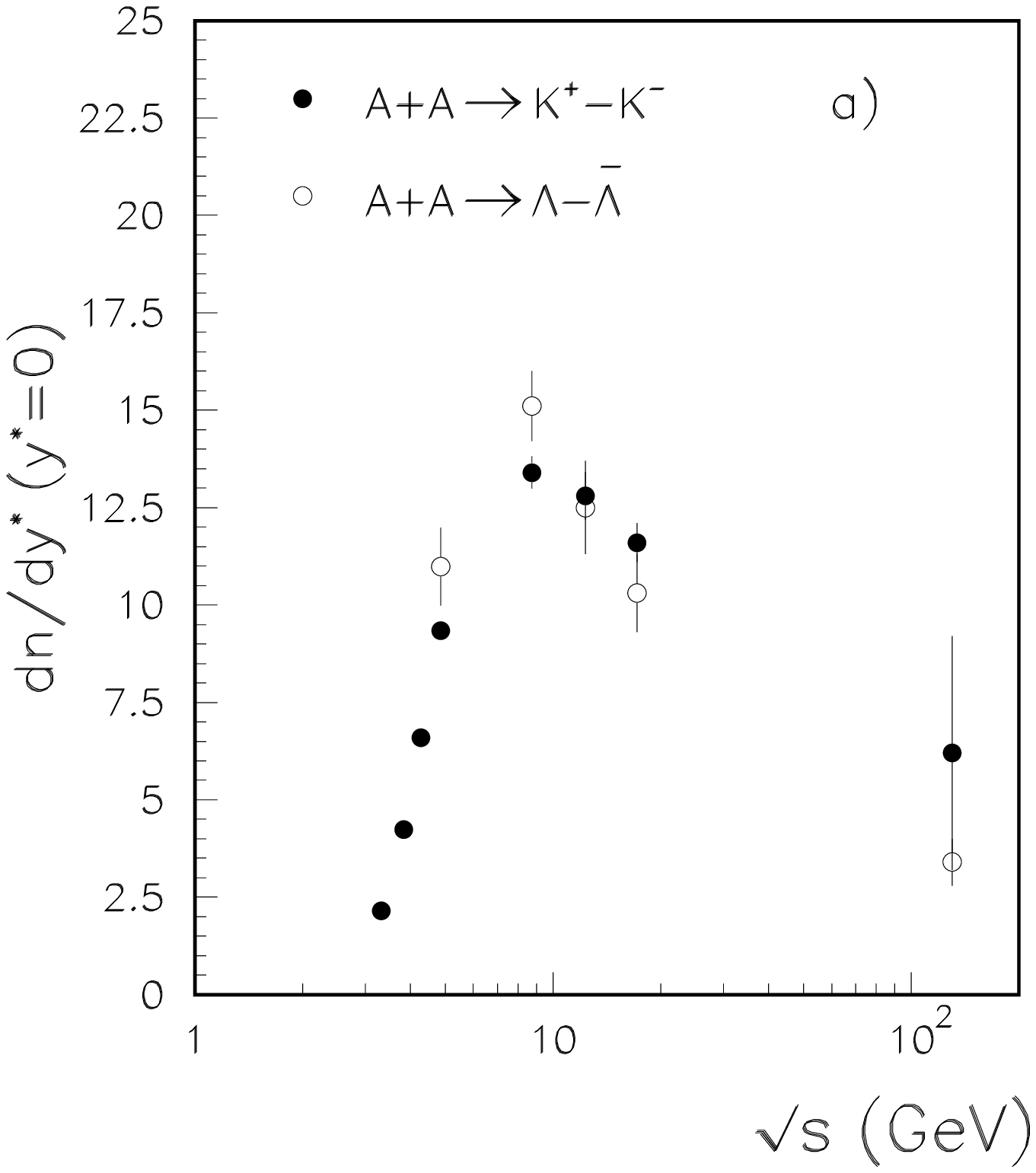,width=8.0cm,height=8.0cm}
\hspace{-2.cm}
\epsfig{file=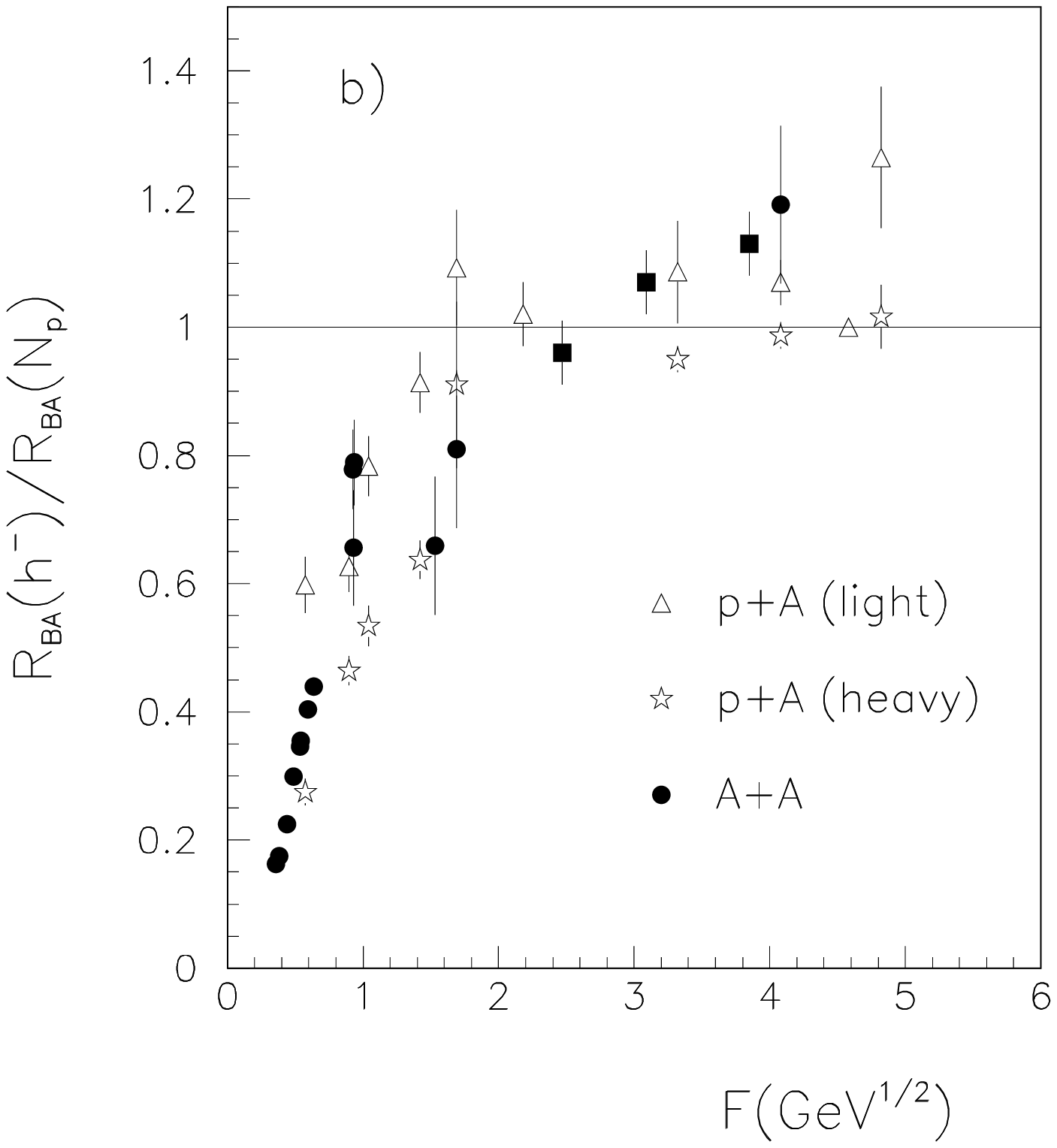,width=8.0cm,height=8.0cm}
\caption{a) Energy dependence of  $\Lambda$$ -$$\bar{\Lambda}$
and $K^{+}$$ -$$K^{-}$
mid-rapidity
multiplicities  b) The ratio $R_{BA}(h^{-})$/ $R_{BA}(N_{p})$  for p+A and A+A reactions versus Fermi energy F (for the
explanation see text)}
\label{fig:toosmall}
\end{figure}

 The energy dependence of $K^{+}$ and $K^{-}$ mid-rapidity multiplicities
in  central A+A collisions
is shown in Fig.~2b ( preliminary NA49 data \cite{kollegger2001} are indicated by squares).
Due to the  associated production of $K^{+}$ with
$\Lambda$ particles at lower energies
 one would expect a similar behaviour of singly produced $\Lambda$ baryons and
singly produced $K^{+}$ mesons. The results on the energy dependence of the $K^{+}$ $-$ $K^{-}$ 
and  $\Lambda$ $-$ $\bar{\Lambda}$ yields at mid-rapidity, shown in Fig.~3a, confirm this
expectation. We want to emphasize that the energy dependence of the $\Lambda$ $-$
$\bar{\Lambda}$ yield at mid-rapidity in p+p (see Fig.~1a) shows a similar pattern.
Unfortunately, the  corresponding  $K^{+}$ and $K^{-}$ data at lower energies are still missing. 

 Before we proceed with the results on  the strangeness to pion multiplicity ratios, we would 
like to comment on the energy dependence of pion multiplicities observed in A+A and p+A collisions.    
In order to study this dependence, we define the double ratio of multiplicities per wounded nucleon
\begin{equation}
\frac{R_{BA}(h^{-})}{R_{BA}(N_{p})} =
\frac{\langle h^{-}\rangle _{BA}}{\langle h^{-}\rangle _{NN}}/
\frac{\langle N_{W}\rangle _{BA}}{\langle N_{W}\rangle _{NN}}
\end{equation} 
where $\langle h^{-}\rangle _{BA}$ ($\langle h^{-}\rangle _{NN}$) is the mean number of negative particles
and $\langle N_{W}\rangle _{BA}$ ($\langle N_{W}\rangle _{NN}$) is   
the mean number of wounded nucleons in B+A (N+N) collisions. Fig.~3b shows the dependence of the double 
ratio $R_{BA}(h^{-})$/ $R_{BA}(N_{p})$ on the Fermi energy 
F=$(\sqrt{s} -2\cdot m_{p})^{\frac{3}{4}})/\sqrt{s}^{\frac{1}{4}}$ for the protons on light (A$\le$64) 
(triangles)  and
heavy (A$\ge$108) (stars) nuclei, together with  the results from A+A collisions (circles) \cite{marek}.
 Preliminary NA49 data are indicated by squares. In the Wounded-Nucleon-Model (WNM)  \cite{wnm} it is
assumed
that the
number of
produced particles depends only
on the total number of wounded nucleons $N_{W}$ participating in the collisions, and not on the number of
collisions per participating nucleon. Under this assumption the ratio  $R_{BA}(h^{-})$/ $R_{BA}(N_{p})$ $=$ 1.   
The A+A results from Fig.~3b  show that the WNM overpredicts the $h^{-}$ yield at lower
energies  ($R_{BA}(h^{-})$/ $R_{BA}(N_{p})$ $<$ 1) and
underpredicts at higher energies ($R_{BA}(h^{-})$/ $R_{BA}(N_{p})$ $>$ 1). Before we interpret this
$h^{-}$ enhancement, relative to the WNM, as an entropy enhancement and a 
phase transition indication  a similar behaviour in p+A has to be critically considered (see
Fig.~3b).
It is interesting to note that the  observed $R_{BA}(h^{-})$/ $R_{BA}(N_{p})$ ratio is larger for 
the p
reactions on light nuclei than on heavy  nuclei. This may indicate that the WNM overestimates the
number of the
participating nucleons in the proton collisions on  heavy nuclei.
\begin{figure}[]
\vspace{-1.cm}
\epsfig{file=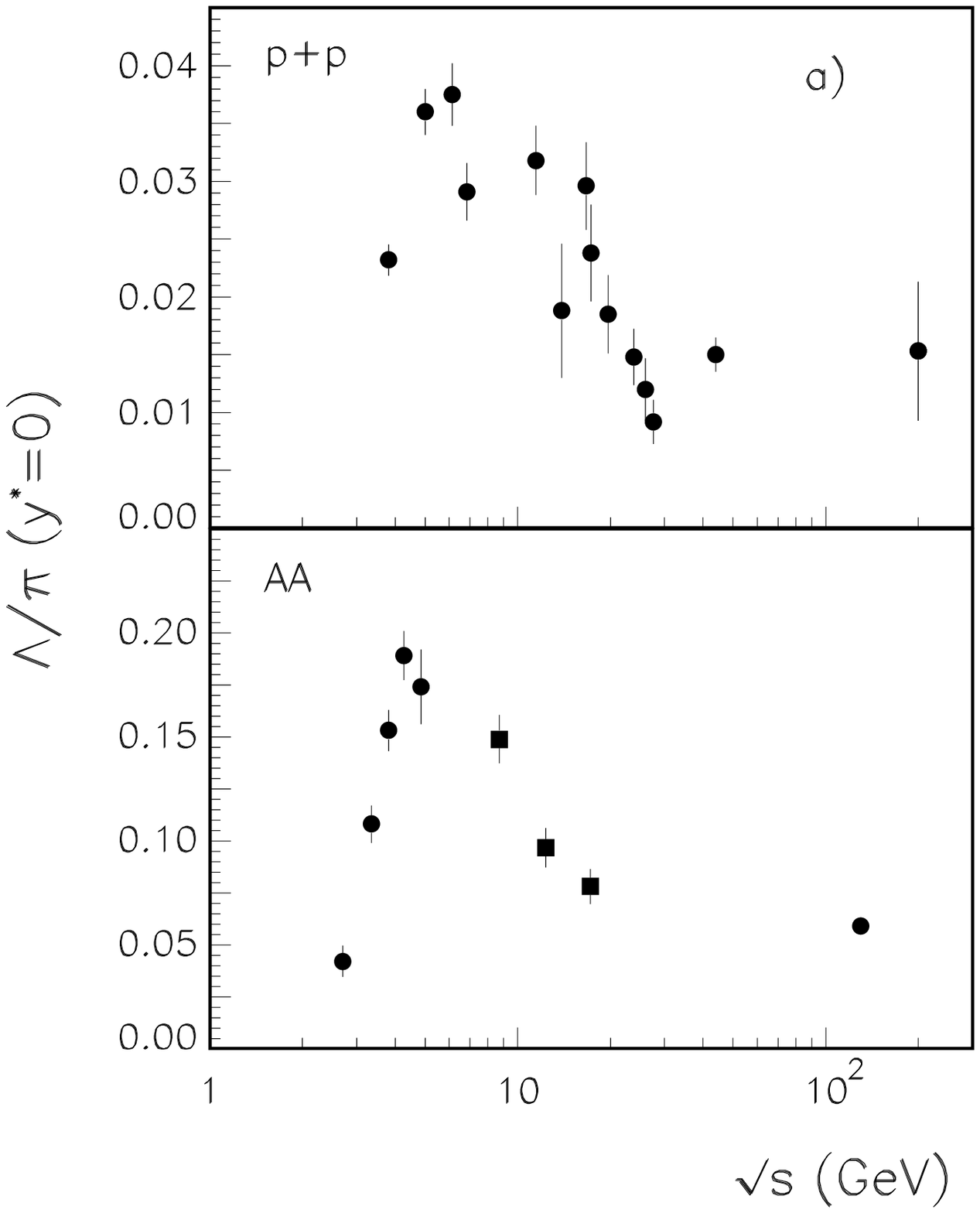,width=9.0cm,height=10.0cm}
\hspace{-2.8cm}
\epsfig{file=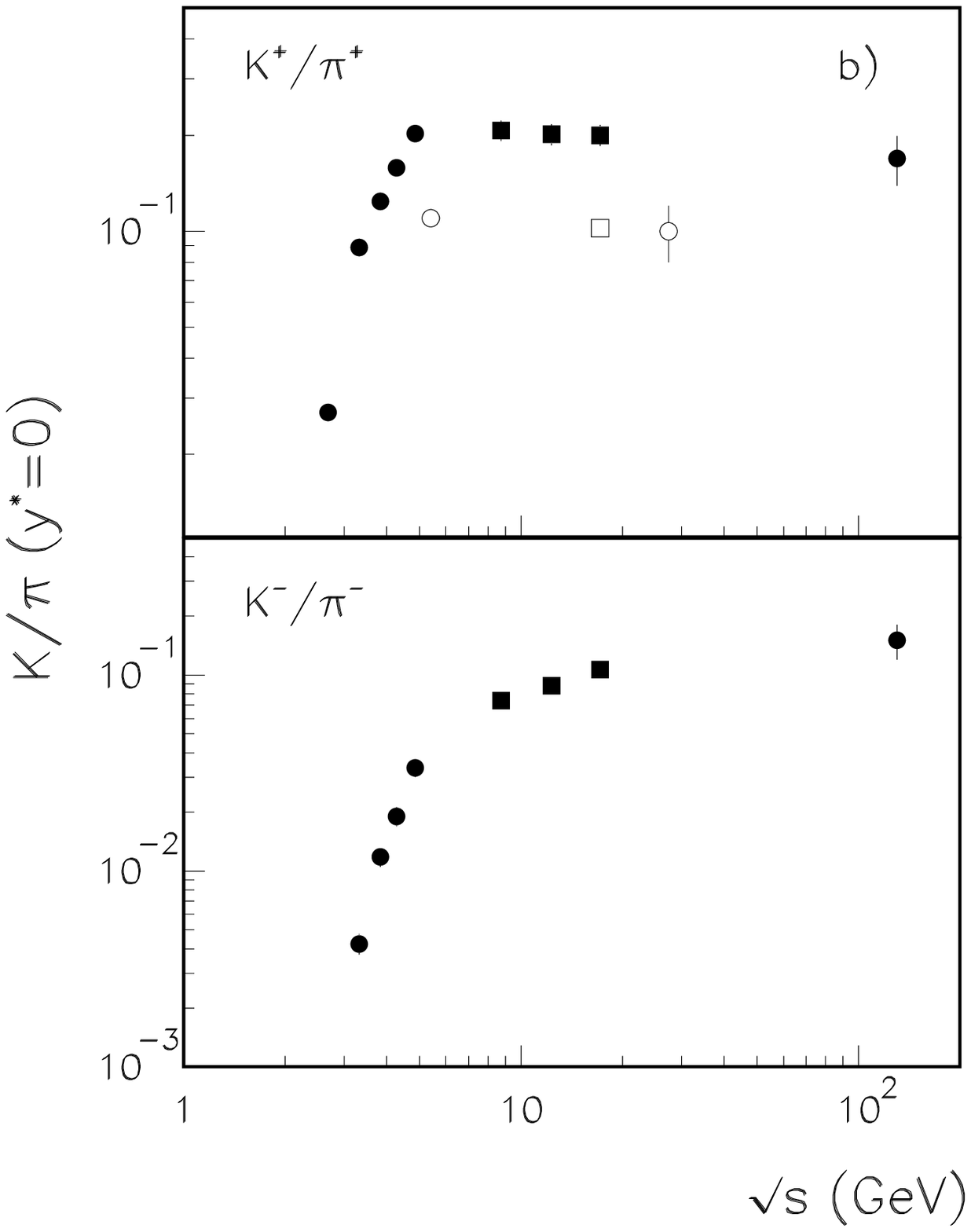,width=9.0cm,height=10.0cm}
\caption{a) Energy dependence of $\Lambda$/$\pi$ ratio at mid-rapidity in p+p (upper plot) and A+A (lower
plot).
b) Energy dependence of $K^{+}$/$\pi^{+}$ (upper plot) and $K^{-}$/$\pi^{-}$ (lower plot) at mid-rapidity in
A+A (full symbols) and p+A \cite{pa} (open symbols). Preliminary NA49 results are
indicated  by squares.}
\label{fig:toosmall}
\end{figure}
  Because of the similarity in the production of $\Lambda$ particles at mid-rapidity in p+p and
A+A collisions, it is not
surprising that
the energy dependence of the $\Lambda$/$\pi$ ratio at mid-rapidity in p+p
and  A+A shows a similar behaviour: non-monotonic, passing through a
maximum  around $\sqrt{s}$=4$-$7 GeV (Fig.~4a).

 The p+A and A+A data show (upper part of Fig.~4b) that the  $K^{+}$/$\pi^{+}$ ratio is  affected
similarly, due to the  associated production of $K^{+}$ together with
$\Lambda$
particles at lower energies, where the net-baryon density is
still high.  While a continuous rise in $K^{-}$$/$$\pi^{-}$ from AGS over
SPS to RHIC energies is observed
 in A+A data 
(lower part of Fig.~4b), the $K^{+}$$/$$\pi^{+}$ ratio in p+A and A+A data  reaches a maximum 
approximately at the energy where the maximum in the $\Lambda$$/$$\pi$ ratio is observed.
Consequently, it may be premature to conclude that a rapid change of the  energy
dependence of strangeness
to pion (entropy) ratio is a unique signal for  the transition from
confined to
deconfined matter.  From this point of view the study  of
baryon and anti-baryon production in  p+p and p+A collisions, starting
with the low AGS energies, and going up to the top RHIC energy,
will be extremely important for a  better
understanding  of  baryon propagation in rapidity and/or $x_F$ space  in
A+A interactions.   
\begin{figure}[]
\vspace{-1.cm}
\hspace{1.cm}
\epsfig{file=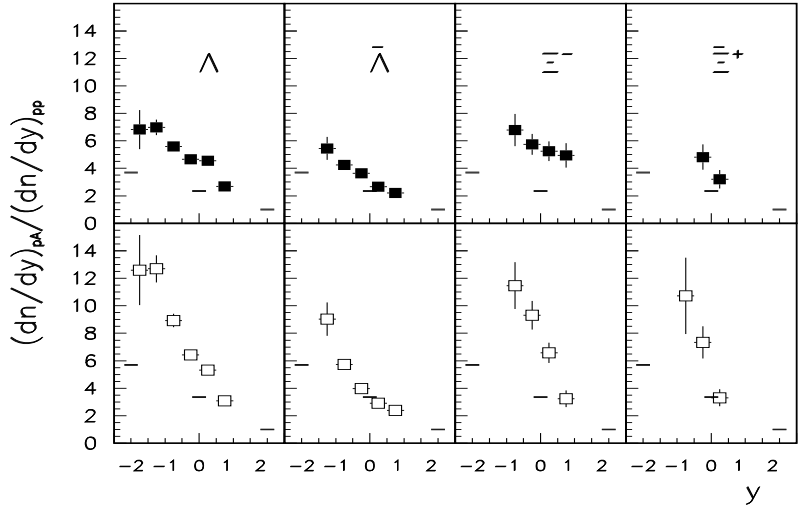,width=11.cm,height=9.cm}
\vspace{-0.5cm}
\caption{Ratios of hyperon multiplicities in centrality selected p+Pb and minimum bias p+p collisions.
The experimental results for the first ($\langle \nu \rangle$$=$ 3.7) and the second
 ($\langle \nu \rangle$$=$ 5.7) centrality bin, are  shown by  full and open squares, respectively.
  The predictions of the WNM at  backward, central and forward rapidities are indicated by the
 solid lines.}
\label{figure1}
\end{figure}

  In addition to  the similarity  between p+p and A+A data, the energy
 dependence of the $\Lambda$/$\pi$ ratio (Fig.~4a) also shows an important
 difference. The $\Lambda$/$\pi$ ratio in A+A collisions is several times 
 larger than the corresponding ratio in minimum bias p+p collisions.
 The  WA97 experiment \cite{wa9799} has found  that the strange particle
yields
 ($\Lambda$,  $\bar{\Lambda}$,  $\Xi^{-}$, $\bar{\Xi}^{+}$ and
 $\Omega^{-}$+
 $\bar{\Omega}^{+}$) per wounded nucleon $N_{W}$ at central rapidity increase from p+Pb to Pb+Pb. 
 The enhancement, relative to the prediction of the WNM
 is found to be more pronounced for multistrange particles, and
 more pronounced for particles than for anti-particles.
  The relevant question is: does the simple WNM scaling hold between
 p+p and p+A interactions?

 Fig.~5 shows the  ratios of hyperon multiplicities  produced in centrality selected p+Pb and 
 minimum bias p+p collisions  versus  rapidity \cite{tanja2001}.  The centrality  of p+Pb
 reactions is characterized by the mean number of projectile collisions
$\nu$= $N_{W}$$-$1.
 The experimental results for the first ($\langle \nu \rangle$$=$ 3.7) and the second 
 ($\langle \nu \rangle$$=$ 5.7) centrality bin are  shown by  full and open squares, respectively. 
  The predictions of the WNM at backward ($\frac{(dn/dy)_{pA}}{(dn/dy)_{pp}}$$=$$\nu$), central 
 ($\frac{(dn/dy)_{pA}}{(dn/dy)_{pp}}$$=$$\frac{\nu+1}{2}$) and forward
($\frac{(dn/dy)_{pA}}{(dn/dy)_{pp}}$$=$$1$)
rapidities are
indicated by solid lines.  
   As already observed in Pb+Pb collisions \cite{wa9799}, the mid-rapidity enhancement in p+Pb, relative to the WNM
 prediction, 
is larger for particles of higher strangeness content and is larger for
baryons than for antibaryons. Although the yield of $\Xi^{-}$,
 $\bar{\Xi}^{+}$ and $\Lambda$ hyperons scales approximately with $\nu$ as
predicted by
the WNM (not shown here),
it is important to emphasize that  their
absolute yield is enhanced.  The yield of  $\bar{\Lambda}$
is close  to the WNM for the most central sample. 

 In the backward rapidity region there is also a clear
indication for the enhanced production of  $\Lambda$,  $\bar{\Lambda}$,
$\Xi^{-}$ and $\bar{\Xi}^{+}$ particles relative to the WNM model,
similar to the effect observed at the mid-rapidity region.

\begin{figure}[] 
\vspace{-1.cm}
\hspace{2.5cm}
\epsfig{file=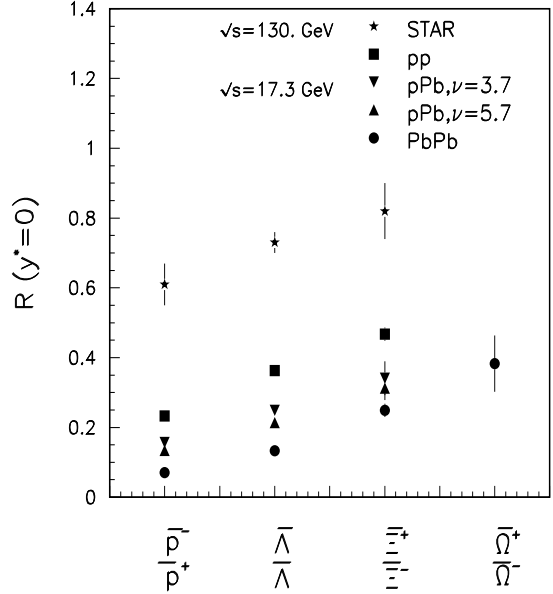,width=9.0cm,height=10.0cm}
\caption{ Antibaryon to baryon  ratios  at mid-rapidity. 
Preliminary NA49 results from   p+p, p+A and A+A at $\sqrt{s}=$17.2 GeV are compared
with  $\frac{\bar{\Omega}}{\Omega}$ ratio  at $\sqrt{s}=$17.2 \cite{wa9799} and the
STAR results \cite{star2001}}
\label{fig:toosmall}
\end{figure}   
 The observed enhancement of strange and multistrange baryons in p+Pb
 suggests  that any extrapolation from elementary to nucleus-nucleus
collisions using the simple WNM is questionable, as well as the
conclusions based on the result of this extrapolation.
 Before one can conclude
that the enhancement with respect to the WNM seen in A+A is a sign for
QGP
formation \cite{wa9799}, one has to study carefully the consequences of
the multiple collision mechanisms as they become accesible in p+A
interactions. 

A short summary of presented data, to emphasize  again the importance of  energy and system
size dependence study, is shown in Fig.~6 in the form of antibaryon to baryon ratios at mid-rapidity 
for p+p, centrality selected p+Pb, central Pb+Pb at $\sqrt{s}=$17.2 GeV,
as well as  central Au+Au at
$\sqrt{s}=$130 GeV. 
  The "slope" of the antibaryon to baryon ratios at a given energy becomes
flatter (going from
$\frac{\bar{p}}{p}$ to 
$\frac{\bar{\Xi}}{\Xi}$ or $\frac{\bar{\Omega}}{\Omega}$) 
as the size of the colliding
system (going from A+A to p+p), and consequently  the net-baryon density at mid-rapidity,
decreases. A similar effect is obtained by increasing the  energy and keeping  
(approximately) the same system  size (compare particle ratios from the central Pb+Pb and Au+Au  collisions
at SPS and RHIC energies, respectively). The ratios of $\frac{K^{-}}{K^{+}}$ will
be affected in a similar way
since  the production of the $K^{+}$ mesons is strongly correlated with change of the net-baryon
density, as it was discussed before. It would be very interesting to compare the antibaryon to baryon
ratios    
in  p+p and  A+A at the different energies at which they  have  approximately the same  $\frac{K^{-}}{K^{+}}$
ratio. In doing so the isospin effect should be carefully taken into account.
\section{Conclusions}
 The NA49 Collaboration at the CERN SPS has
 measured strange hyperons in
minimum bias p+p and centrality selected p+Pb reactions at 158 GeV/c, and 
$\Lambda$, $\bar{\Lambda}$, $K^{+}$ and $K^{-}$ particles  in central Pb+Pb collisions at
40, 80 and 158 A$\cdot$GeV.  
 A steeper $p_T$ slope is observed in p+p reactions for $\bar{\Lambda}$s (pair
produced $\Lambda$s)  than for the
 $\Lambda$s (pair and non-pair produced $\Lambda$s)
reflecting  their  different production
mechanisms.
It should be critically considered  how this
 matches the physics assumptions of the statistical model regardless
of its success in describing particle yields in p+p reactions.   
 The same question concerns the applicability of the statistical
model
in describing the particle yields in the fragmentation region of A+A
reactions.

 By comparing the energy dependence of $\Lambda$ and $\bar{\Lambda}$
production at mid-rapidity, a striking similarity is observed between p+p
and A+A data. This is also seen in the energy dependence of the
$\Lambda$/$\pi$
ratio $-$ non-monotonic, passing through a maximum around $\sqrt{s}$$=$4$-$7
GeV.  The $K^{+}$/$\pi$ ratio at mid-rapidity is  affected similarly in p+A and A+A 
 interactions due to the  associated production of $K^{+}$ together with
$\Lambda$ particles, in particular  at the lower energies where the
net-baryon density is
still relatively high. From this we conclude that a rapid change of
energy dependence
of strangeness to pion (entropy) ratio may not be a unique characteristic
of heavy-ion reactions.

  Comparing the yields of produced hyperons in p+p and p+Pb reactions an
excess of $\Lambda$,
$\bar{\Lambda}$, $\Xi^{-}$ and $\bar{\Xi}^{+}$ is observed
relative to the prediction of the Wounded-Nucleon-Model. This excess
shows a similar pattern to the one observed in Pb+Pb data, indicating  that the extrapolation
to A+A using the WNM is questionable, as well as the conclusions based on this
extrapolation.

 We  would like to stress that the observed similarity
between p+p,
p+A and A+A reactions does not prove that A+A is a simple superposition  
of p+p and/or p+A. It shows however, that a deeper
understanding of p+p and p+A data is necessary for any
interpretation of  A+A reactions.

\end{document}